\documentclass[preprint,12pt]{aastex}

\shorttitle{Detector GRB Sensitivity}

\begin{document}

\title{Comparison of the Gamma-Ray Burst Sensitivity of Different Detectors}
\author{David L. Band\altaffilmark{1}}
\affil{GLAST SSC, Code 661, NASA/Goddard Space Flight Center,
Greenbelt, MD  20771}
\altaffiltext{1}{Joint Center for Astrophysics, Physics
Department, University of Maryland Baltimore County,
1000 Hilltop Circle, Baltimore, MD 21250}
\email{dband@lheapop.gsfc.nasa.gov}

\begin{abstract}
Gamma-ray burst detectors are sensitive at different energies,
complicating the comparison of the burst populations that they
detect.  The instrument teams often report their detector
sensitivities in their instruments' energy band.  I propose that
sensitivities be reported as the threshold peak photon flux $F_T$
over the 1--1000~keV energy band for a specific spectral shape.
The primary spectral parameter is $E_p$, the energy of the maximum
$E^2 N_E\propto \nu f_\nu$.  Thus $F_T$ vs. $E_p$ is a useful
description of a detector's sensitivity.  I find that Swift will
be marginally more sensitive than BATSE for $E_p>100$~keV, but
significantly more sensitive for $E_p<100$~keV.  Because of its
low energy sensitivity, the FREGATE on {\it HETE-2} is
surprisingly sensitive below $E_p=100$~keV. Both the WFC on {\it
BeppoSAX} and the WXM on {\it HETE-2} are/were sensitive for low
$E_p$. As expected, the GBM on {\it GLAST} will be less sensitive
than BATSE, while {\it EXIST} will be significantly more sensitive
than Swift.  The {\it BeppoSAX} GRBM was less sensitive that the
WFC, particularly at low $E_p$.
\end{abstract}


\section{Introduction}
The gamma-ray burst missions that have flown in the past 15 years
have reported that bursts are hard (e.g., BATSE---Preece et al.
2000; Mallozzi et al. 1995) or soft (e.g., Ginga---Strohmayer et
al. 1998).  However these missions had different sensitivities to
hard and soft bursts, and have reported the threshold burst
intensities using a variety of different measures, such as peak
flux or energy fluence measured over different energy bands. To
synthesize the results of these mission (and determine whether
they are mutually consistent), and to compare the capabilities of
past, current, and proposed missions, we need common measures of
burst intensity and hardness, and we need to express a burst
detector's capabilities in terms of those common measures.

I propose to characterize bursts by the peak photon flux
integrated over 1--1000~keV averaged over one second, and by the
spectrum's peak energy $E_p$ during this one second. The peak
energy $E_p$ is the energy of the maximum of $N(E)E^2$
(proportional to $\nu F_\nu$)---the energy flux per logarithmic
energy (frequency) band; $E_p$ is a first order measure of the
spectral hardness.  The choice of peak photon flux (as opposed to
energy fluence, energy flux, or total photon fluence) is based not
on the physics of bursts (i.e., theories of fundamental burst
properties) but on detector triggers:  most detectors trigger on a
statistically significant increase in the count rate over a
specified energy band.  The detection threshold is the minimum
peak count rate for which the detector would have triggered.  Even
though detectors trigger on the count rate in different energy
bands, the peak count rate can be translated into the peak photon
flux over a fiducial energy band with knowledge of the burst
spectrum and the detector's energy response.  Note that the count
rate is a detector-dependent quantity while the photon flux is an
intrinsic description of the burst photons arriving in the Solar
System. Therefore, the peak photon flux over a common energy band
provides a convenient instrument-independent measure of burst
intensity.

Because few bursts have lightcurves where the maximum flux
is constant over seconds, the peak photon flux will depend
on the accumulation time $\Delta t$, the time resolution
with which the flux is measured. Although most detectors
trigger on a variety of accumulation times, $\Delta t=1$~s
is usually included in a detector's set of accumulation
times.  In addition, a detector's sensitivity for one value
of $\Delta t$ can be translated into the average
sensitivity for other values using an ensemble of burst
lightcurves (Band 2002).  Therefore I use $\Delta t=1$~s
for this work.

Many studies have presented results using peak photon fluxes in
the 50--300~keV band, principally because this was main trigger
band of the Burst and Transient Source Experiment (BATSE) on the
{\it Compton Gamma-Ray Observatory (CGRO)}.  While any fiducial
energy band can be used, particularly soft transient events, such
as the recently-discovered X-Ray Flashes (XRFs, which may or may
not be related to classical gamma-ray bursts---Heise et al. 2001),
will produce little flux in the 50--300~keV band. Therefore I
choose to use the 1--1000~keV band because most burst detectors
operate within this broad band, and most bursts have $E_p$ in this
band.

This study focuses predominantly on a comparison of the
sensitivity of a number of detectors.  In \S 2 I discuss
the methodology used in this comparison, while \S 3
presents the relevant information about each detector.  The
results and their implications are discussed in \S 4.

\section{Methodology}

I assume that the burst spectrum can be described by the
``GRB'' function (Band et al. 1993),
\begin{eqnarray}
N(E) =& N_0 \left({E\over \hbox{100 keV}} \right)^\alpha
   \exp\left[-{E\over{E_0}}\right] \hfil & ; \quad E \le E_b \nonumber \\
   & N_0 \left({E_b\over \hbox{100 keV}} \right)^{\alpha-\beta}
   \exp\left[\beta-\alpha \right]
   \left({E\over \hbox{100 keV}} \right)^\beta \hfil
   & ; \quad E > E_b \\
   \hbox{where} & E_b = (\alpha-\beta) E_0 \hfil
    & \hbox{and } \quad E_p = (\alpha+2) E_0 \quad .\nonumber
\end{eqnarray}
$N(E)$ has units of photons~s$^{-1}$~keV$^{-1}$~cm$^{-2}$.
I compare the sensitivity of different detectors by holding
$\alpha$ and $\beta$ fixed, and varying $E_p$. To compare
spectra with different values of $E_p$ I use the integral
of $N(E)$ over a broad energy band
\begin{equation}
F = \int_{E_l}^{E_h} N(E) dE
\end{equation}
where $E_l=1$~keV and $E_h=1000$~keV.  Fig.~1 compares the
fluxes in the 50--300~keV and 1--1000~keV bands as a
function of $E_p$ for different sets of $\alpha$ and
$\beta$.  As can be seen, the 1--1000~keV flux includes
more photons for small $E_p$.

The sensitivity of a detector to a burst with a particular
spectral shape depends on its burst trigger and hardware
properties such as the detector area $A$, the fraction of the
detector that is active $f_{\rm det}$, the fraction of the coded
mask that is open $f_{\rm mask}$, and the efficiency
$\epsilon(E)$.  If the detector does not have a coded mask, then
$f_{\rm mask}=1$. Most gamma-ray detectors do not have a
one-to-one mapping between a photon's energy and the energy
channel the detector assigns a detected count.  If the incoming
photon spectrum is binned in energy, the relationship between the
photon energy bins and the detector's energy channels is a
detector response matrix (DRM) with off diagonal elements and not
a simple detector efficiency for each energy bin. However, since I
calculate the count rates over broad energy bands, the DRM can be
approximated by a detector efficiency function $\epsilon(E)$.

The typical burst trigger looks for a statistically
significant increase in the detector's count rate above the
background in the energy band between $E_1$ and $E_2$ in a
time bin $\Delta t$.  The significance of the increase is
measured in units of the expected fluctuation scale of the
background, i.e., the square root of the expected number of
background counts.  The count rate increase is assumed to
result from the burst flux. In most cases I model the
background (counts~s$^{-1}$~keV$^{-1}$~cm$^{-2}$) as
\begin{equation}
B(E) = \epsilon(E) \Omega f_{\rm mask} N_B(E)+B_{\rm int}
\end{equation}
where $N_B(E)$ is the diffuse high energy background (Gruber
1992), $\Omega$ is the average solid angle of the sky as seen from
the detector plane (calculated from the corrected formulae in
Sullivan 1971), $f_{\rm mask}$ is the fraction of the coded mask
that is open, and $B_{\rm int}$ is the internal background.  Note
that the aperture flux resulting from the diffuse background is
detected with the detector's efficiency $\epsilon(E)$. This
background model is clearly a simplification because it does not
attempt to model explicitly the background induced by the particle
flux, scattering off the spacecraft and the Earth's atmosphere,
and activation of the detector and its environs; many of these
effects are included in $B_{\rm int}$. At higher energies (e.g.,
$\sim 100$~keV) the instrument walls may become transparent, and
$\Omega$ may increase; this effect is not considered here.
Nonetheless, this model gives an approximate magnitude and energy
dependence. In some cases I use the observed background rates.

The trigger has a preset threshold significance
\begin{equation}
\sigma_0 = {{A f_{\rm det} f_{\rm mask} \Delta t \int_{E_1}^{E_2}
   \epsilon(E) N_T(E)dE}\over
   \sqrt{ A f_{\rm det} \Delta t \int_{E_1}^{E_2} B(E) dE}}
\end{equation}
where $N_T(E)$ is the peak burst flux at the threshold, and
$f_{\rm det}$ is the fraction of the detector plane that is active. In
eq.~4 the numerator is the number of counts from the burst and the
denominator is the square root of the number of counts expected
from the background rate. Therefore, the broadband photon flux at
the detector's threshold is
\begin{equation}
F_T = {{\int_{E_l}^{E_h} N_T(E) dE}\over
   {\int_{E_1}^{E_2} \epsilon(E) N_T(E) dE}}
   {{\sigma_0\sqrt{\int_{E_1}^{E_2} B(E) dE}}
   \over {f_{\rm mask} \sqrt{A f_{\rm det} \Delta t}}} \quad .
\end{equation}
Note that the ratio of integrals over $N_T$ eliminates the
unknown normalization $N_0$ in eq.~1, and results in the inverse
of the average efficiency. I vary the spectral shape of $N_T(E)$
by varying $E_p$, holding $\alpha$ and $\beta$ fixed. The
resulting $F_T (E_p)$ compares different detectors.

For each detector I need information about the detector and about
the burst trigger.  For the detector I need the area $A$, the
detector efficiency $\epsilon(E)$, the fraction of the detector
that is active $f_{\rm det}$, the fraction of the coded mask that
is open $f_{\rm mask}$, the average solid angle $\Omega$, and the
internal background $B_{\rm int}$; for the trigger I need the
effective threshold significance $\sigma_0$ and the trigger energy
band $E_1$--$E_2$. Most detector papers present a plot of
$\epsilon(E)$; I model $\epsilon(E)$ as a series of power laws
between representative values of $E$.  Detectors use triggers with
a variety of trigger times $\Delta t$ and energy ranges (defined
by $E_1$ and $E_2$). The goal of this study is to compare
detectors with different energy responses and triggering on
different energy bands, and not to study the effect of different
trigger times. Almost all detectors include $\Delta t$=1~s among
their set of trigger times, and this will be the value used here.

In the next section I present the detectors in this study; Table~1
summarizes some of the detector parameters. I use a detector's
maximum sensitivity, even if achieved over only a small region
within the field-of-view (FOV).
\section{The Detectors}
\subsection{{\it CGRO}'s BATSE}
BATSE consisted of eight modules, each with two types of
detectors:  the Large Area Detector (LAD) for burst detection,
localization and monitoring; and the Spectroscopy Detector (SD)
for spectral analysis.  Thus the LADs are relevant to this study.
The LADs were built around large (2025 cm$^{2}$), flat NaI(Tl)
crystals.  The LADs in the eight modules were parallel to the
faces of a regular octahedron. The LAD effective area curve is
found in Fishman et al. (1989). The LADs operated in the
20--2000~keV band, but usually triggered in the 50--300~keV band.

BATSE triggered as a whole when two or more LADs each
triggered on an increase greater than $\sigma_0=5.5$.
Therefore the sensitivity depended on the second most
brightly illuminated detector for which the cosine of the
angle to the source (the factor by which the flux is
diminished) varied between 1/3 (when the burst was along
the normal to the most brightly illuminated detector) and
0.8165 (when the burst was exactly between two detectors).
Note that this analysis ignores the effects of scattering
off of the spacecraft and the Earth's atmosphere. In this
study I use the maximum sensitivity in the detector's FOV.
Thus for BATSE the required significance is equivalent to
$\sigma_0=5.5/0.8165=6.74$.
\subsection{{\it BeppoSAX}'s WFC}
{\it BeppoSAX}'s Wide Field Cameras (WFCs) were two anti-parallel
coded mask detectors which pointed perpendicular to the axis of
the Narrow Field Instruments (Jager et al. 1997).  The detector
plane was a 25.5$\times$25.5~cm$^2$ multi-wire proportional
counter that was active over 0.8 of its area.  Only 1/3 of the
1~mm$^2$ mask pixels were open, but because of the supports for
the mask pixels, the actual open area of an open pixel was
$0.9\times0.9$~mm$^2$. Because the mask and the detector plane
were almost exactly the same dimension, the spatial sensitivity
was triangular; I present the sensitivities at the center.

The WFC did not trigger on-board, but instead the rates in 1~s and
8~s time bins were analyzed on the ground. Subsequently the rates
accumulated over 1, 5 and 20~minutes were also searched for
transients.  A $\sigma_0=4$ increase in the count rate from one
WFC unit triggered further analysis: the set of time bins with the
highest signal-to-noise ratio were used to create an image, and a
point source with a 5.5$\sigma$ significance was required to
consider the burst real (J.~Heise, personal communication, 2002).
Thus for the analysis here $\sigma_0=4$.
\subsection{{\it BeppoSAX}'s GRBM}
The Gamma-Ray Burst Monitor on {\it BeppoSAX} consisted of
the four 1136 cm$^2$ area, 1~cm thick CsI(Na) shields
around the Phoswich Detection System (Feroci et al. 1997;
Amati 1999); detecting bursts was the secondary role of
these shields. For most of the mission the system triggered
when the 40--700~keV rate in two detectors accumulated over
$\Delta t=1$~s increased by more than $\sigma_0=4$ (Feroci
et al. 1997; Amati 1999).

The average background count rate in the trigger band for
each detector was $\sim 1000$~cts/s, and I use this rate
rather than model the background count rate.  The effective
areas of the four detectors differed because of material
that was in front of them, but I model them as having been
identical.  Because the count rate must have increased by
$\sigma_0=4$ in two detectors, the most sensitive point in
the GRBM's FOV for the on-board trigger was between the
normal to two adjoining detectors, that is, at an angle of
45$\arcdeg$ from each detector normal.  Thus the effective
significance was $\sigma_0=4\sqrt{2}=5.66$.

The ultimate GRBM burst database is the result of a search on the
ground with a variety of trigger criteria (Guidorzi 2001)
utilizing the rates in the 40--700~keV and the $>100$~keV bands
from different sets of detectors.  The background was estimated
either as a constant rate calculated from count rates before the
burst or as a linear fit to count rates before and after the
burst.  These complicated trigger criteria might lower the
sensitivity curve by $\sim1/3$. However, in the figure I show only
the sensitivity for the on-board trigger.
\subsection{{\it HETE-2}'s WXM}
The Wide-field X-ray Monitor (WXM) is the primary detector on the
{\it High Energy Transient Explorer~II (HETE-2)} for the
localization of gamma-ray bursts (Kawai et al. 2002). The WXM
consists of two coded mask X-ray detector units sensitive to the
burst's position in one dimension; the units' orientations are
orthogonal to each other, providing a two-dimensional position.
Each WXM unit has a coded mask 18.7~cm above two
position-sensitive proportional counters (PSPCs).  The mask is 1/3
open.  The geometric area of each PSPC is $8.35 \times
12.$~cm$^2$, and they are separated by a gap of 1~cm. The
detectors operate over the 2--25~keV band.  I use the background
count rate of $\sim 700$ for both detectors provided by Kawai et
al. (2002).  A variety of triggers are used, with different
trigger significances; I use $\sigma_0=5.9$ for the $\Delta t=1$~s
accumulation in the rate summed over both detectors. A burst must
be imaged after a rate trigger, effectively raising the
significance for the detection of a burst, but I do not attempt to
model this effect.
\subsection{{\it HETE-2}'s FREGATE}
The FREGATE (Atteia et al. 2002) is a set of 4 NaI(Tl) detectors
on {\it HETE-2}.  The FREGATE's goals are a)~the detection of
bursts for localization by the imaging cameras, b)~burst
spectroscopy and c)~monitoring hard X-ray transient sources.  Each
detector has a circular area of 39.6~cm$^2$. The active area is
exposed to the sky without a coded mask, and the FREGATE has no
localization capabilities.  The shield around each NaI crystal
extends 2.7~cm above the front surface of the crystal (J.-L.
Atteia, private communication, 2003), reducing both the FOV and
the aperture flux.  The detector has a diameter of 7.1~cm, while
the shield has an inner diameter of 8.0~cm.  The detectors are
sensitive in the 6--400~keV band, and triggers on the 6--40~keV,
6--80~keV, 32--400~keV, and $>400$~keV count rates.  The rates
from one set of two detectors are combined into a summed rate, and
the rates from the other set of two detectors are combined into a
second summed rate.  False triggers are eliminated by requiring a
rate increase of 4.5$\sigma$ in both summed rates.
\subsection{Swift's BAT}
The Burst Alert Telescope (BAT) is Swift's gamma-ray
instrument. The BAT will detect the gamma-ray burst,
localize it, and instruct the spacecraft to slew so that
the burst location is in the much smaller FOV of the X-ray
and optical telescopes coaligned with the BAT.  The BAT
will use a two step trigger: first, it will detect an
increase in the count rate, and second, it will image the
burst. Only if the count rate increase originates from a
point source will the event be considered an astrophysical
transient. Here I consider only the count rate trigger.

The BAT will consist of a rectangular detector plane of 32,768 CZT
detectors, each 4$\times$4~mm$^2$; because of the packaging of the
detectors, the total active area of 5243~cm$^2$ is spread over
$\sim7200$~cm$^2$. A D-shaped coded mask with a total area of
3.2~m$^2$ and 5$\times$5~mm$^2$ cells will be one meter above the
detector plane.  I use an efficiency curve provided by
C.~Markwardt (personal communication, 2002).

My formula for the average solid angle $\Omega$ seen by the
detector plane is for a rectangular mask above a rectangular
detector plane; the BAT has a more complicated geometry. To
calculate $\Omega$ I assume the mask is 120$\times$250~cm$^2$ and
the detector plane is 60$\times$120~cm$^2$.  Only a fraction $\sim
0.72$ of this detector plane is active, but the dimensions of the
region over which the active area is spread are required. Note
that $\Omega$ is used to calculate the background rate, and the
threshold flux is proportional to the square root of the
background; therefore the result will not be very sensitive to
small errors in $\Omega$ resulting from this approximation.

The Swift rate trigger will be very flexible, utilizing many
different energy bands, background estimates, and accumulation
times (E.~Fenimore, personal communication, 2002). These different
triggers will use different significances. To compare the
sensitivity as a function of energy of different detectors, I
consider only $\Delta t$=1~s, and I use the energy bands currently
planned (see Table~1; D.~Palmer, personal communication, 2002).
Although the rate triggers for $\Delta t$=1~s may first trigger on
$\sigma_0 \sim 6$, the requirement that a new source appear in the
image of the source image effectively raises $\sigma_0$ to
$\sim8$.  Note that the BAT will trigger not only on the rate from
the entire detector plane, but also on the rate from subsets,
increasing the sensitivity in the partially-coded FOV
(E.~Fenimore, personal communication, 2002). The maximum
sensitivity will be at the center of the fully-coded region normal
to the BAT.
\subsection{{\it GLAST}'s GBM}
The Gamma-ray Burst Monitor (GBM) planned for the {\it Gamma-ray
Large Area Space Telescope (GLAST)} will consist of 12 NaI
detectors to cover the 5--1000~keV band and two BGO detectors to
cover the 1--30~MeV band.  The purpose of the GBM is to detect
bursts in or near the FOV of the Large Area Telescope (LAT), {\it
GLAST}'s main instrument, and to characterize the bursts. Since
the LAT is a high energy gamma-ray detector, the sensitivity to
particularly hard bursts is relevant. Here I focus on the
sensitivity of the NaI detectors, which will be built around flat
127~cm$^2$ NaI(Tl) crystals that will each view $\sim$half the sky
(von Kienlin et al. 2000). The two BGO detectors will provide
spectral coverage between the NaI detectors and the LAT, but will
not be useful for detecting bursts (as I have verified with this
paper's methodology).

Included among the variety of burst triggers for the NaI detectors
will be BATSE-like triggers where two or more detectors must
trigger. The orientation of the 12 detectors has not been
finalized, but the smallest angle between two detectors will be of
order $\sim30\arcdeg$, and thus the most sensitive points in the
FOV will have an angle to the second most brightly illuminated
detector of $\sim15\arcdeg$. Since $\cos(15\arcdeg)\sim 1$ I use
the threshold significance of $\sigma_0=4.5$.  The trigger will
use the standard BATSE energy band of 50--300~keV, as well as
other energy bands.  By experimenting with a variety of energy
bands I find that the sensitivity is maximized for 10--100 and
50--300~keV; these energy bands are used in Figure~8.
\subsection{\it EXIST}
The {\it Energetic X-ray Imaging Survey Telescope (EXIST)} is
currently proposed to be a free-flying mission to detect gamma-ray
bursts and conduct a hard X-ray sky survey (Grindlay et al. 2002).
{\it EXIST} will carry three identical telescopes, each of which
will consist of 9 coded mask modules.  The modules' detector
planes will be canted with respect to each other by
10--12.5\arcdeg.  A module's detector plane will be 3000~cm$^2$ of
CZT in front of a CsI anticoincidence shield.  The CZT detectors
will be 5--10~mm thick, increasing the high energy efficiency
relative to Swift; I use the efficiency curve for 5~mm thick CZT
in my calculations. Additional CsI planes will form 90~cm high
collimator walls between the modules. The active CsI shielding can
also be used as a high energy gamma-ray burst detector, but its
effects will not be considered here.  A curved coded mask will
arch over all of a telescope's 9 modules 150~cm above the detector
planes.

To determine a telescope's sensitivity I calculate the sensitivity
for a single module, and then consider how the modules'
sensitivity add together over the sky for a single telescope.  The
fully-coded regions of all 9 modules in a telescope do not
overlap. The maximum sensitivity for the current design is 1.92
times the sensitivity of a single module.
\section{Discussion}
\subsection{Detector Comparison}
The maximum sensitivities for the different detectors are
presented in Figures~2--9; the fraction of a detector's FOV at or
near the maximum sensitivity varies from detector to detector. The
sensitivity decreases away from the detector's normal because of
the projection of the detector plane to the burst (proportional to
the cosine of the inclination angle).  In the partially coded
region of a coded mask detector the detector's walls shadow part
of the detector plane.  In addition, the background induced by the
particle flux, which is modelled crudely in my study, varies over
an orbit, raising and lowering the sensitivity. Thus the curves in
Figures~2--9 will shift up and down (mostly up) for different
angles to the detector and for different parts of the spacecraft's
orbit.

Nonetheless, the curves are a measure of the relative
sensitivities of the different detectors.  The GBM's NaI detectors
will be less sensitive than the BATSE LADs for bursts with
$E_p>100$~keV because they will have much less area, although the
GBM's low energy efficiency will give it a comparable sensitivity
for bursts with low $E_p$. With significantly less area that the
BATSE LADs, the FREGATE is less sensitive for bursts with
$E_p>70$~keV. However, the FREGATE's low energy sensitivity (down
to 6~keV) increases its sensitivity to detect particularly soft
bursts relative to BATSE. Also, FREGATE triggers on a variety of
$\Delta E$ while BATSE triggered on only one value of $\Delta E$.
{\it EXIST} will be more sensitive than Swift because the aperture
flux (per detector area) is almost a factor of 2 smaller while the
detector area is larger. Note that although the total detector
area of just one of {\it EXIST}'s three telescopes will be
27,000~cm$^2$, no point in the FOV is in the fully coded region of
all 9 modules. These relative sensitivities are also affected by
the trigger significance required of a rate increase.

The energy dependence of the detector efficiency affects
the energy dependence of the sensitivity.  CZT has high
efficiency below $\sim 100$~keV and then decreases, while
NaI's efficiency peaks at $\sim 100$~keV and remains high
until $\sim1$~MeV.  Thus the sensitivity of the BATSE LADs
(which were NaI detectors) and the GBM NaI detectors
decreases significantly for $E_p<100$~keV, while the
sensitivity of the CZT detectors (Swift and {\it EXIST})
decrease much less for low $E_p$.  {\it EXIST} will use
thicker CZT detectors than Swift, which will increase the
high energy efficiency, and thus {\it EXIST}'s high energy
sensitivity is greater than Swift's by more than would be
predicted by the increase in area.

A lower low energy cutoff also increases the low energy
sensitivity since more of the spectrum can be detected. This
explains the comparable low energy sensitivities of FREGATE and
Swift, even though Swift's effective area will be much larger than
FREGATE's: FREGATE is sensitive above 6~keV and Swift's spectrum
will begin at $\sim15$~keV.

As expected, the trigger energy bands also affect the energy
sensitivity.  The cusps evident in the curves for FREGATE, Swift,
GBM and {\it EXIST} result from high ($\sim$50--150~keV) and low
($\sim$15--50~keV) trigger bands.  Figure~10 shows the sensitivity
of the four trigger energy bands proposed for Swift.
\subsection{Implications}
Gamma-ray bursts will populate the $E_p$--$F_T$ plane shown in
Figures 2--9.  Thus the detector sensitivities shown on these
figures show which burst populations the detectors will detect.
FREGATE is more sensitive below $E_p=100$~keV than BATSE,
particularly for bursts without a high energy tail (i.e., for
$\beta<-3$). Thus BATSE did not trigger on X-Ray Flashes (XRFs),
transients with low $E_p$ (many of the XRFs {\it Beppo-SAX}'s WFC
detected are untriggered events in the BATSE data---Kippen et al.
2002), whereas the FREGATE has detected XRFs. Mallozzi et al.
(1995) showed that on average bursts become softer as they become
fainter.  Kippen et al. show that the XRFs detected by {\it
Beppo-SAX} appear to be the low intensity extension of the
Mallozzi et al. trend.

A comparison of the BATSE and Swift sensitivities (Figures~2
and~7) indicates that on the 1~s timescale Swift may not detect
fainter bursts with $E_p>100$~keV, but it will detect soft bursts
(e.g., XRFs) that are a factor of $\sim 2$ fainter than BATSE's
threshold.  Swift will trigger on timescales both shorter and
longer than the 64~ms, 256~ms and 1.024~s timescales on which
BATSE triggered, and thus will detect fainter bursts with certain
types of lightcurves (e.g., short bursts).
\section{Summary}
I have presented a method of comparing the energy sensitivities of
different gamma-ray burst detectors.  Since the emphasis is on the
energy sensitivity, I assume that all detectors trigger on a
$\Delta t=1$~s accumulation time; the sensitivity for different
accumulation times can be estimated from an ensemble of burst
lightcurves (Band 2002).  I propose presenting the intensity of a
burst in a common unit: the peak photon flux integrated over the
1--1000~keV band.  The threshold peak flux $F_T$ can then be
calculated for each detector for a particular spectral shape.  The
peak energy $E_p$ (the energy of the maximum of $E^2N_E\propto \nu
f_\nu$) is the first order measure of a spectrum's hardness.  Thus
a plot of $F_T$ vs. $E_p$ is a useful summary of a detector's
sensitivity.

An application of this methodology shows that the sensitivity of
BATSE and Swift are approximately equal above $E_p\sim100$~keV,
while Swift will be more sensitive for bursts with lower values of
$E_p$.  As expected for smaller NaI detectors, the GBM-NaI system
will be much less sensitive than BATSE was for $E_p>100$~keV.
FREGATE is surprisingly sensitive to soft bursts because it
detects photons down to 6~keV. Finally, with its large area, {\it
EXIST} will be more sensitive than previous detectors.

\acknowledgements

I would like to thank L.~Amati, J.-L.~Atteia, M.~Briggs,
N.~Gehrels, J.~Grindlay, C.~Guidorzi, J.~Heise, J.~Hong, N.~Kawai,
C.~Markwardt, J.~Norris and R.~Preece for discussions and comments
upon this work.

\clearpage

\begin{deluxetable}{l c c c c c c c c}
\tablecolumns{9}
\tablewidth{0pt}
\tablecaption{\label{Table}Detector Parameters}
\tablehead{
&
\colhead{BATSE\tablenotemark{a}} &
\colhead{WFC\tablenotemark{b}} &
\colhead{GRBM\tablenotemark{b}} &
\colhead{WXM\tablenotemark{c}} &
\colhead{FREGATE\tablenotemark{c}} &
\colhead{Swift} &
\colhead{GBM\tablenotemark{d}} &
\colhead{\it EXIST} }
\startdata
$A$\tablenotemark{e} & 2025. & 650. & 1136 & 213.6\tablenotemark{f} & 39.6\tablenotemark{f} &
7200 & 127 & 3000\tablenotemark{g} \\
$f_{\rm det}$\tablenotemark{h} & 1. & 0.8 & 1. & 0.938 & 1. & 0.72 & 1. & 1 \\
$f_{\rm mask}$\tablenotemark{i} & 1. & 0.27 & 1. & 0.33 & 1. & 0.5 & 1. & 0.5 \\
$\Omega$\tablenotemark{j} & $\pi$ & 0.123 & $\pi$ & 0.802 & 1.74 & 1.33 & $\pi$ & 0.704 \\
$\sigma_0$\tablenotemark{k} & 6.74 (5.5) & 4 & 5.66 (4) & 5.9 & 4.5 & 8 & 4.5 & 5 \\
$\Delta E$\tablenotemark{l} & 50--300 & 1.8--28 & 40--700 & 2--25 & 6--40
& 15--30 & 10--100 & 10--70 \\
 & & & & & 6--80 & 15--50 & 50--300 & 40--200 \\
 & & & & & 32--400 & 30--75& & 70--350 \\
 & & & & & $>400$ & 50--150 & & 100--1000\\
\enddata

\tablenotetext{a}{LADs, on {\it CGRO}}
\tablenotetext{b}{On {\it BeppoSAX}}
\tablenotetext{c}{On {\it HETE-2}}
\tablenotetext{d}{NaI(Tl) detector, on {\it GLAST}}
\tablenotetext{e}{Geometric detector area, in cm$^2$}
\tablenotetext{f}{Area of a single detector; sensitivity
calculated for 2 detectors}
\tablenotetext{g}{Area of a single module; sensitivity
is calculated for a telescope of 9 modules. Note that
{\it EXIST} will include 3 telescopes.}
\tablenotetext{h}{Fraction of detector plane that is active.}
\tablenotetext{i}{Fraction of the coded mask that is open.}
\tablenotetext{j}{Average solid angle for the aperture flux}
\tablenotetext{k}{Effective threshold significance,
including the angle between the burst and the detector
normal; the nominal significance is in parentheses}
\tablenotetext{l}{Trigger energy band, in keV; may change for
current and future missions}

\end{deluxetable}

\clearpage

\begin{figure}
\plotone{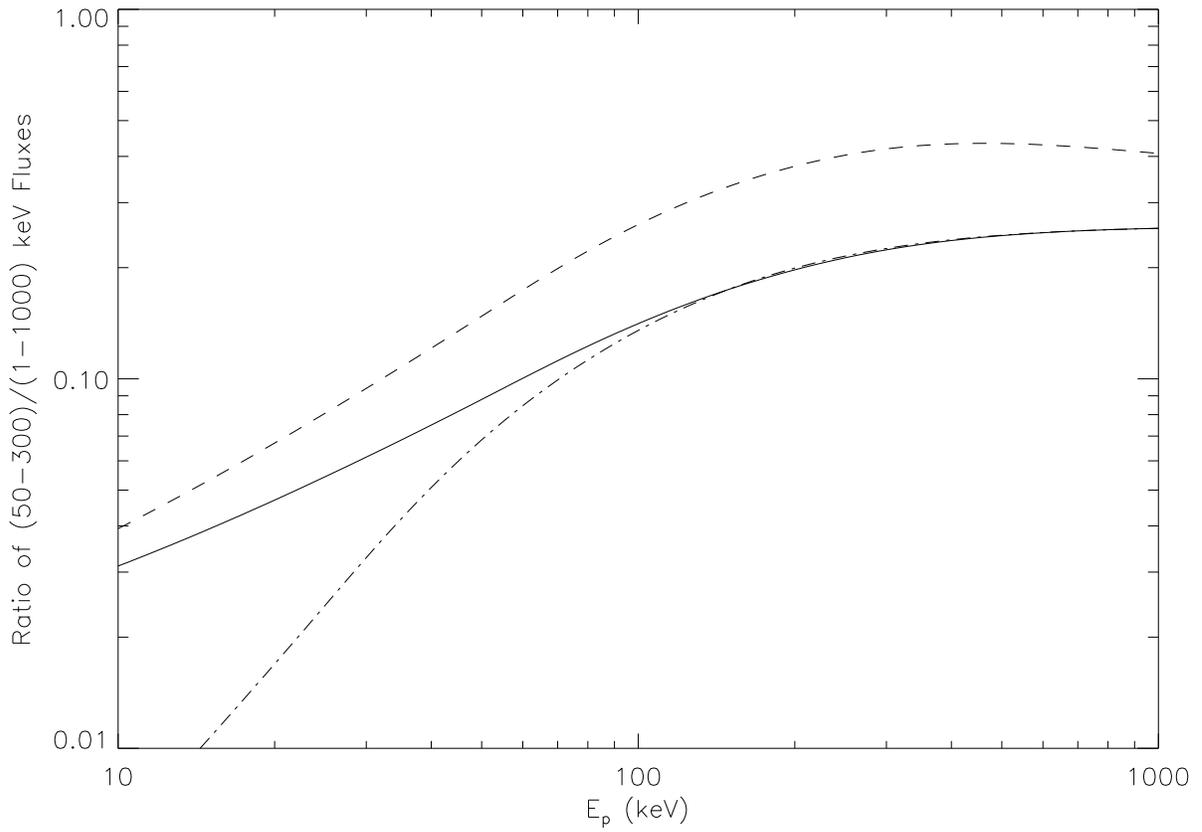}
\caption{Ratio of fluxes in 50--300~keV
to 1--1000~keV bands. Solid line---$\alpha = -1$, $\beta = -2$;
dashed line---$\alpha = -0.5$, $\beta = -2$; dot-dashed
line---$\alpha = -1$, $\beta = -3$. \label{flux_ratio}}
\end{figure}

\begin{figure}
\plotone{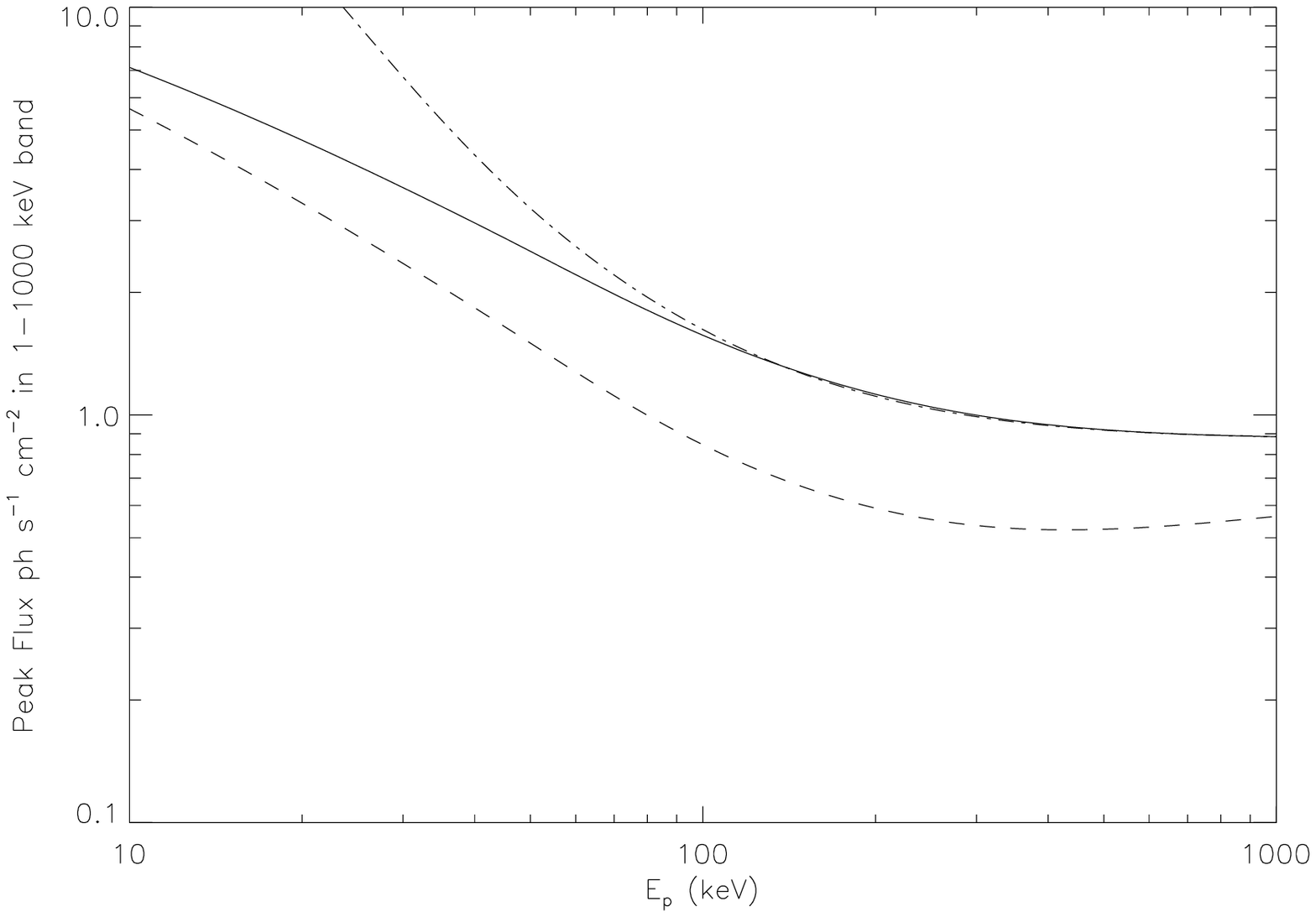}
\caption{Peak flux (1--1000~keV) threshold
of BATSE LAD detectors.  Solid line---$\alpha = -1$, $\beta = -2$;
dashed line---$\alpha = -0.5$, $\beta = -2$; dot-dashed
line---$\alpha = -1$, $\beta = -3$. \label{batse}}
\end{figure}

\begin{figure}
\plotone{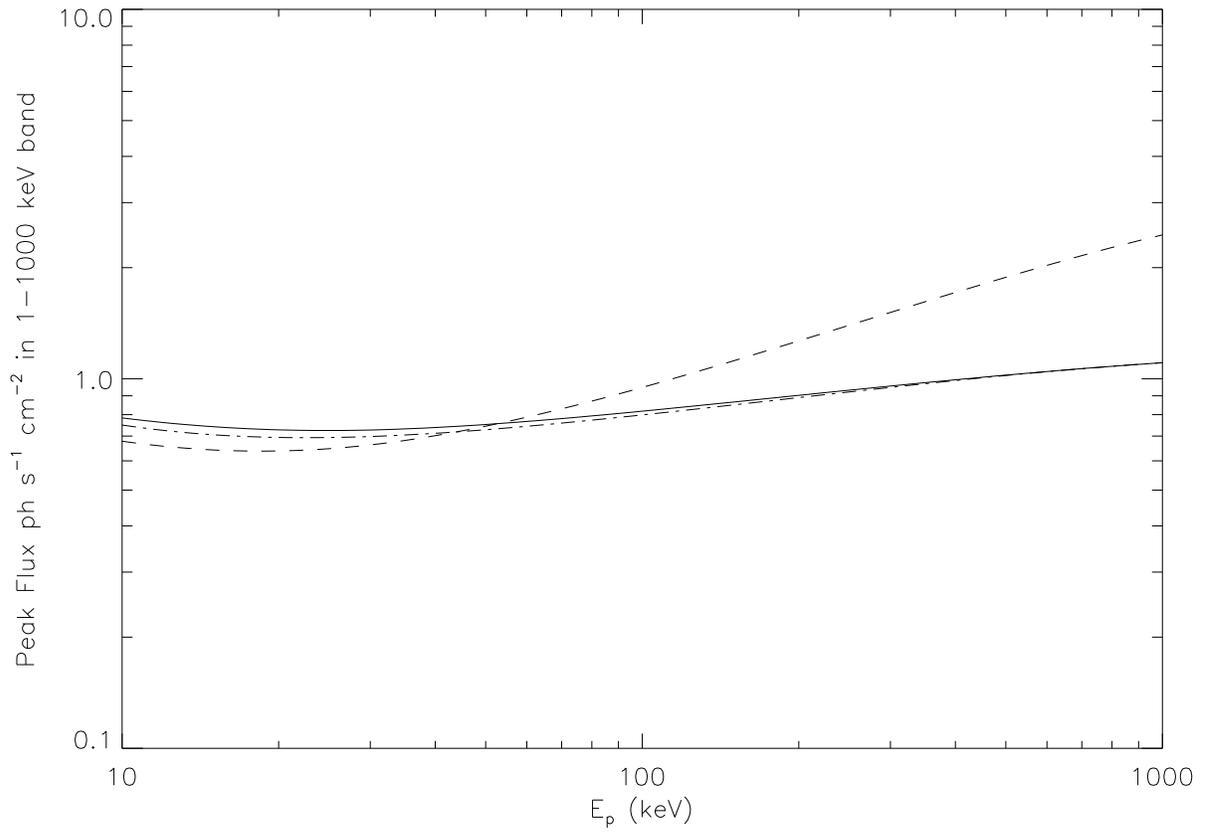}
\caption{Peak flux (1--1000~keV) threshold of the {\it BeppoSAX} WFC.
Solid line---$\alpha
= -1$, $\beta = -2$; dashed line---$\alpha = -0.5$, $\beta = -2$;
dot-dashed line---$\alpha = -1$, $\beta = -3$. \label{wfc}}
\end{figure}

\begin{figure}
\plotone{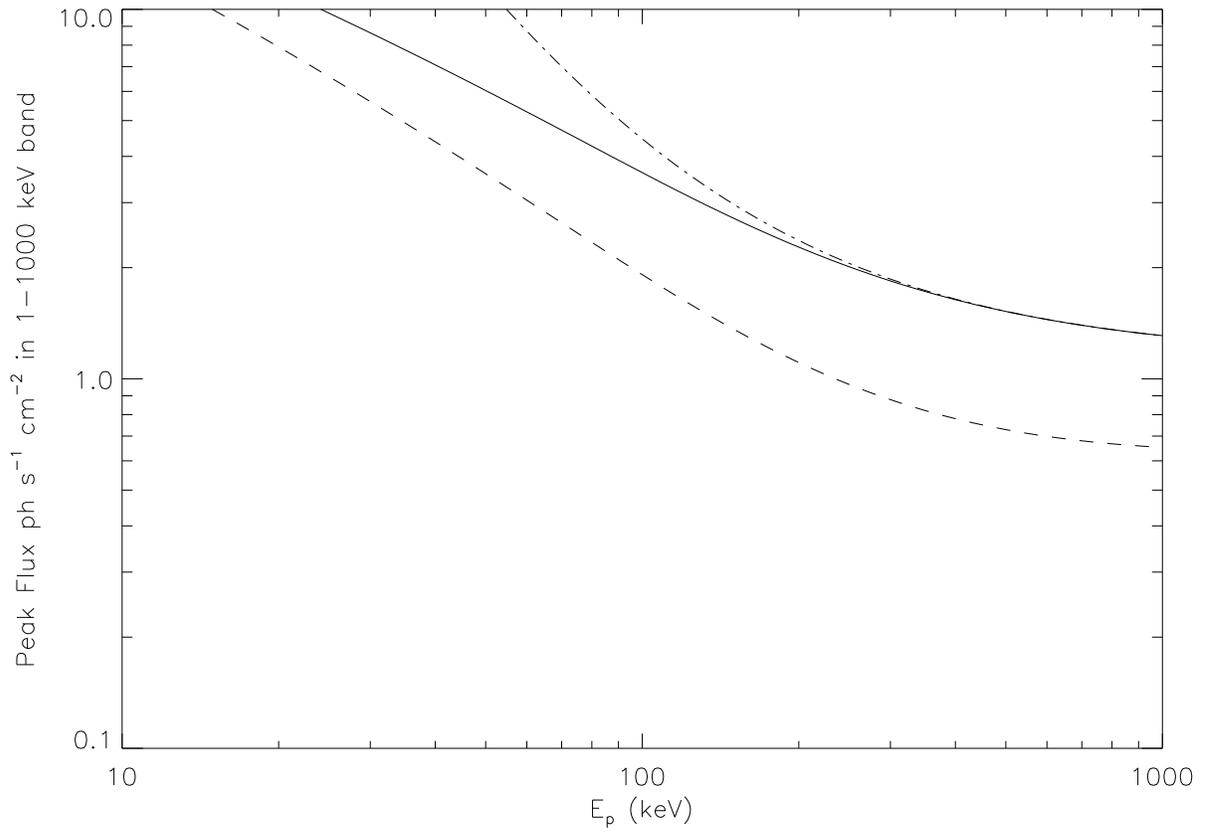}
\caption{Peak flux (1--1000~keV) threshold of the {\it
BeppoSAX} GRBM.  The sensitivity for the on-board trigger
is shown.  Solid line---$\alpha = -1$, $\beta = -2$; dashed
line---$\alpha = -0.5$, $\beta = -2$; dot-dashed line---$\alpha =
-1$, $\beta = -3$. \label{grbm}}
\end{figure}

\begin{figure}
\plotone{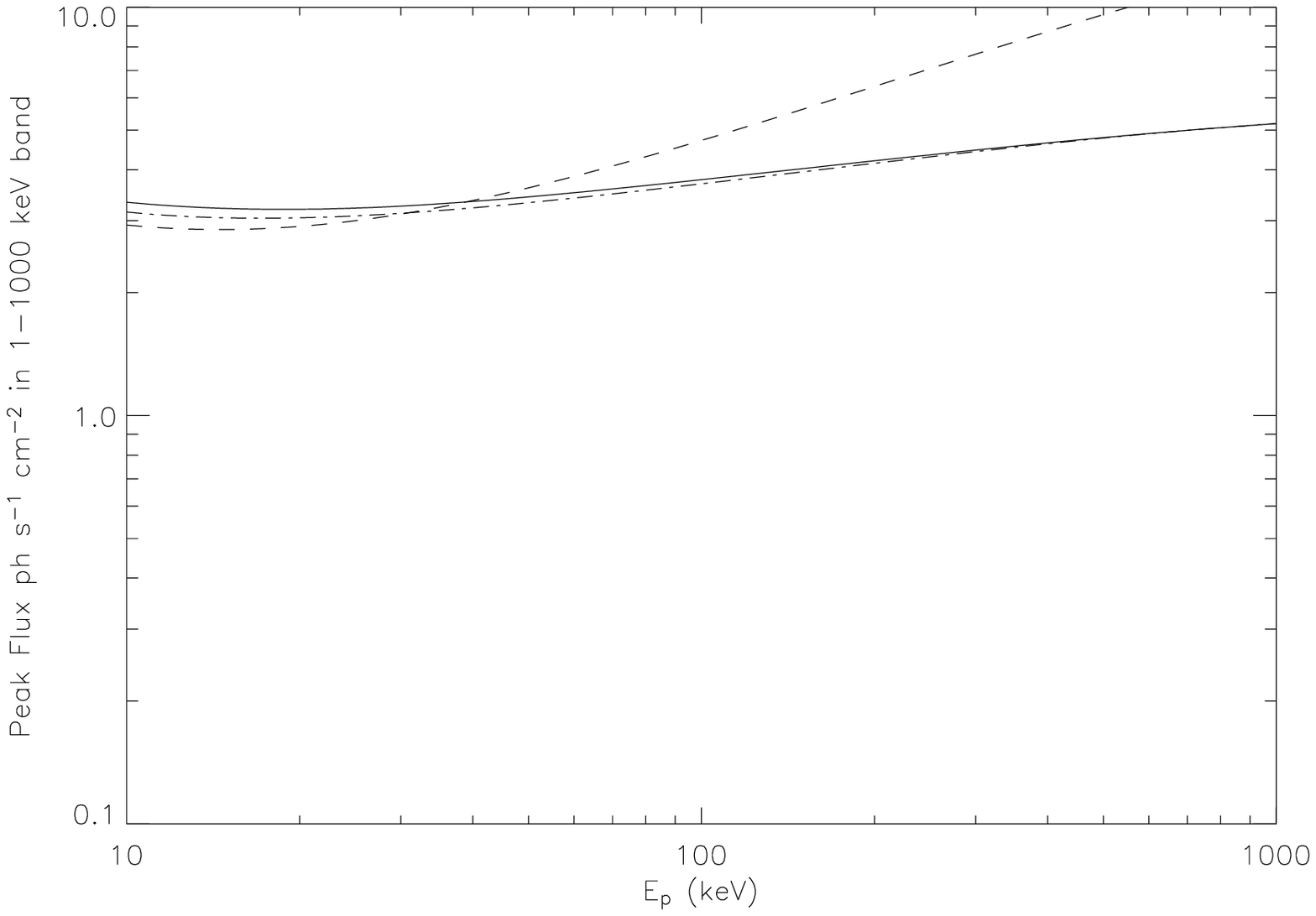}
\caption{Peak flux (1--1000~keV) threshold of the WXM
detectors.  Solid line---$\alpha = -1$, $\beta = -2$;
dashed line---$\alpha = -0.5$, $\beta = -2$; dot-dashed
line---$\alpha = -1$, $\beta = -3$. \label{wxm}}
\end{figure}

\begin{figure}
\plotone{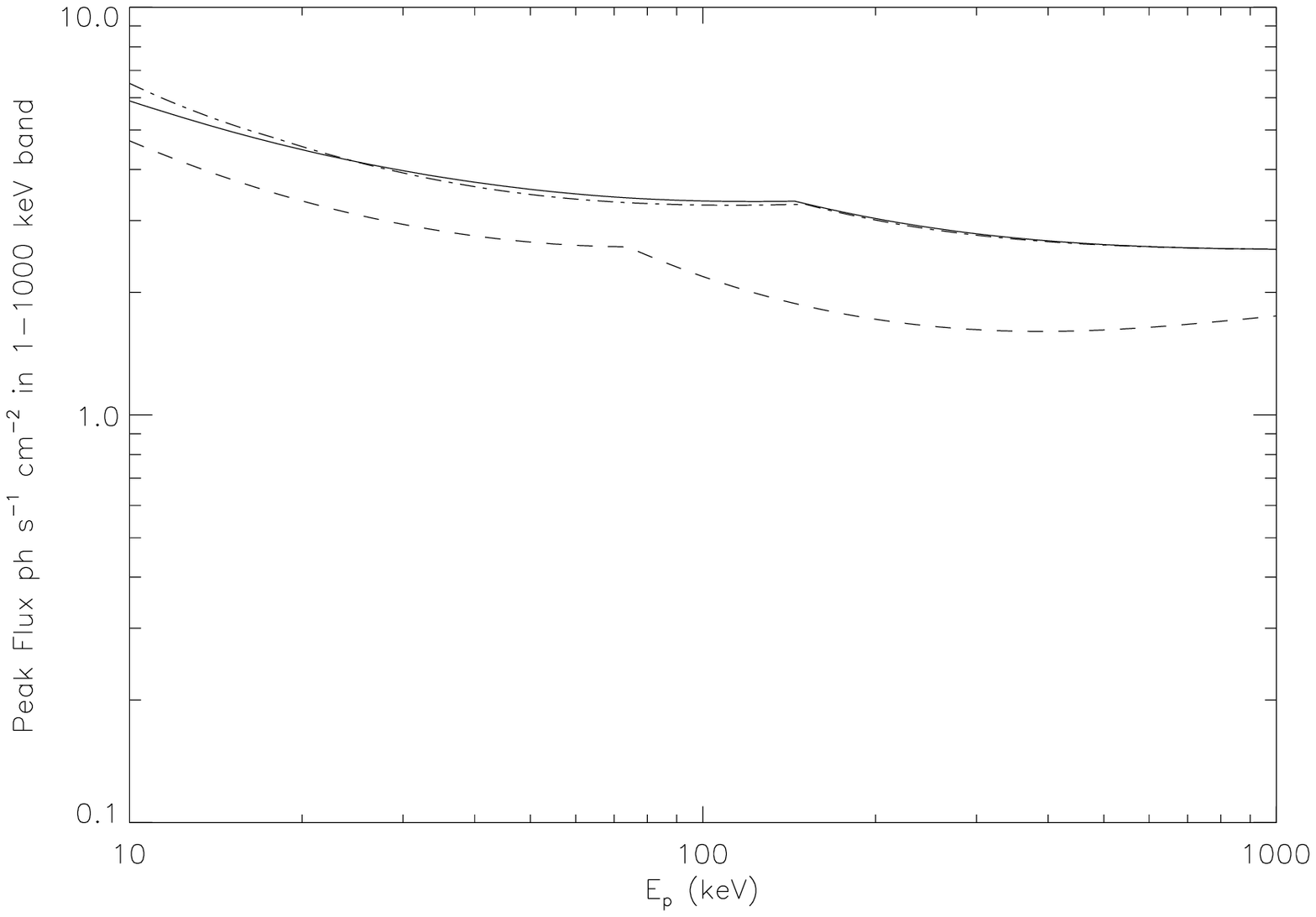}
\caption{Peak flux (1--1000~keV)
threshold of the FREGATE detectors.  Solid line---$\alpha =
-1$, $\beta = -2$; dashed line---$\alpha = -0.5$, $\beta =
-2$; dot-dashed line---$\alpha = -1$, $\beta = -3$.
\label{fregate}}
\end{figure}

\begin{figure}
\plotone{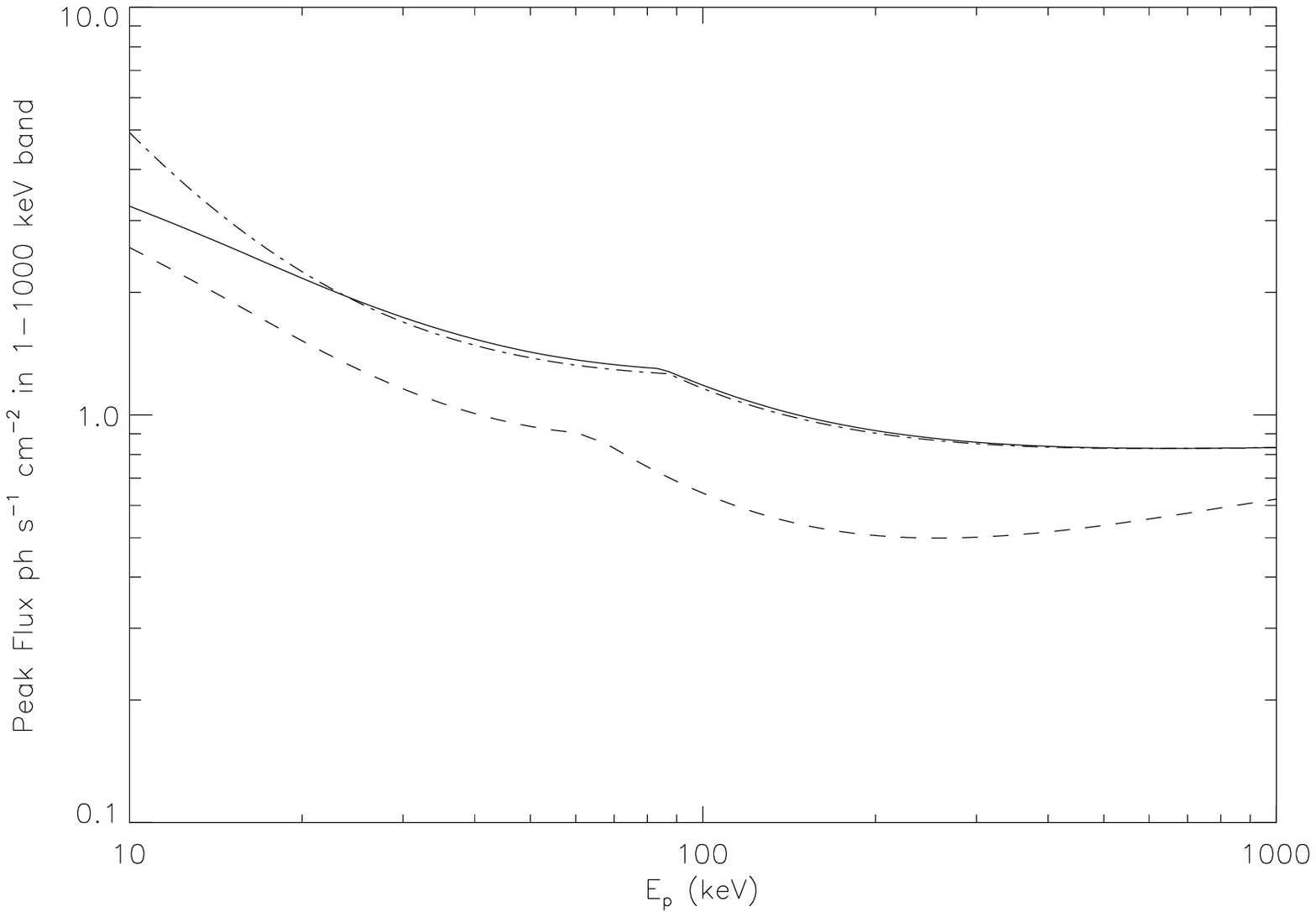}
\caption{Peak flux (1--1000~keV) threshold
of the Swift BAT detector.  Solid line---$\alpha = -1$, $\beta = -2$;
dashed line---$\alpha = -0.5$, $\beta = -2$; dot-dashed
line---$\alpha = -1$, $\beta = -3$. \label{swift}}
\end{figure}

\begin{figure}
\plotone{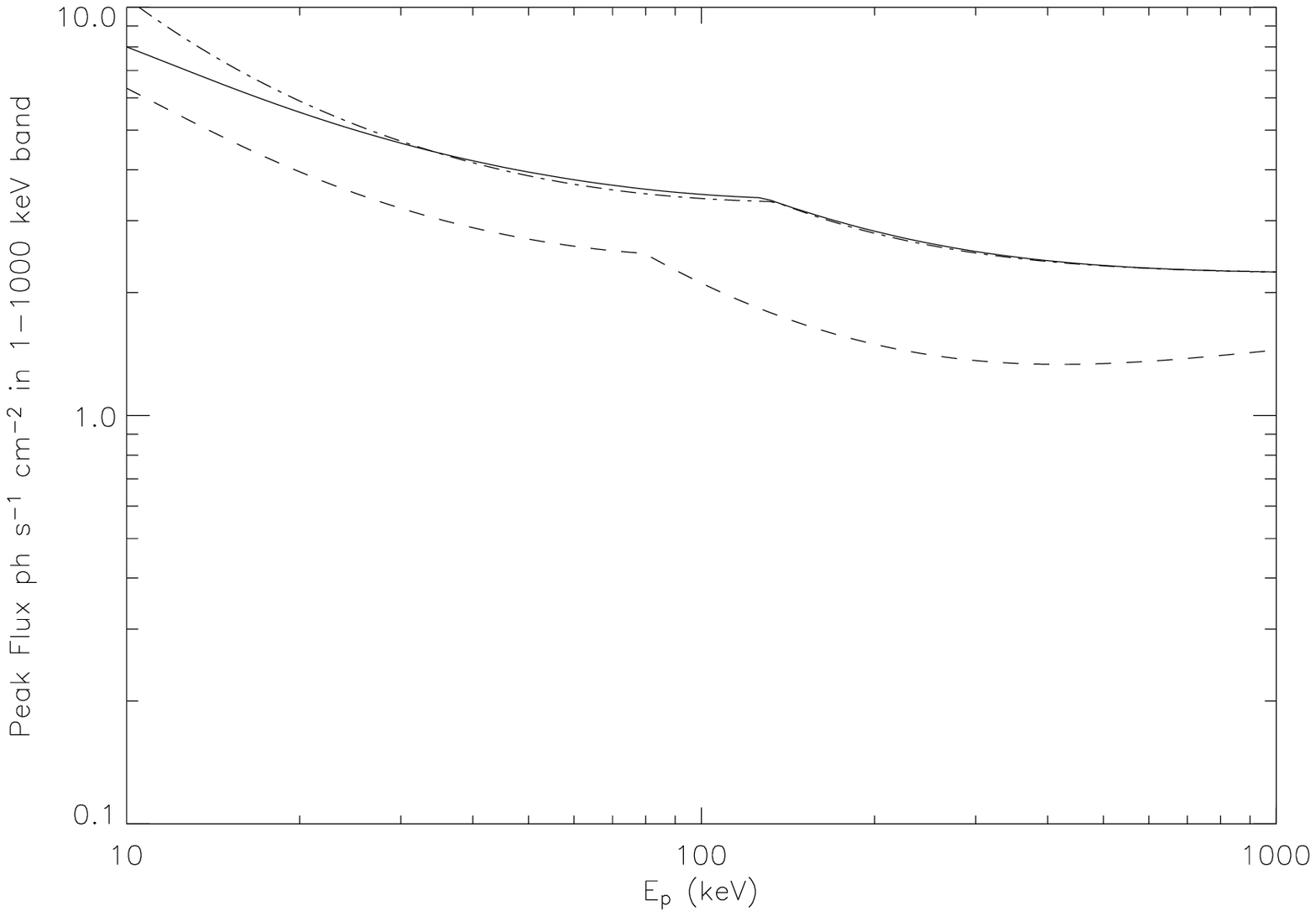}
\caption{Peak flux (1--1000~keV) threshold
of GBM NaI detectors.  Solid line---$\alpha = -1$, $\beta = -2$;
dashed line---$\alpha = -0.5$, $\beta = -2$; dot-dashed
line---$\alpha = -1$, $\beta = -3$. \label{GBM_NaI}}
\end{figure}

\begin{figure}
\plotone{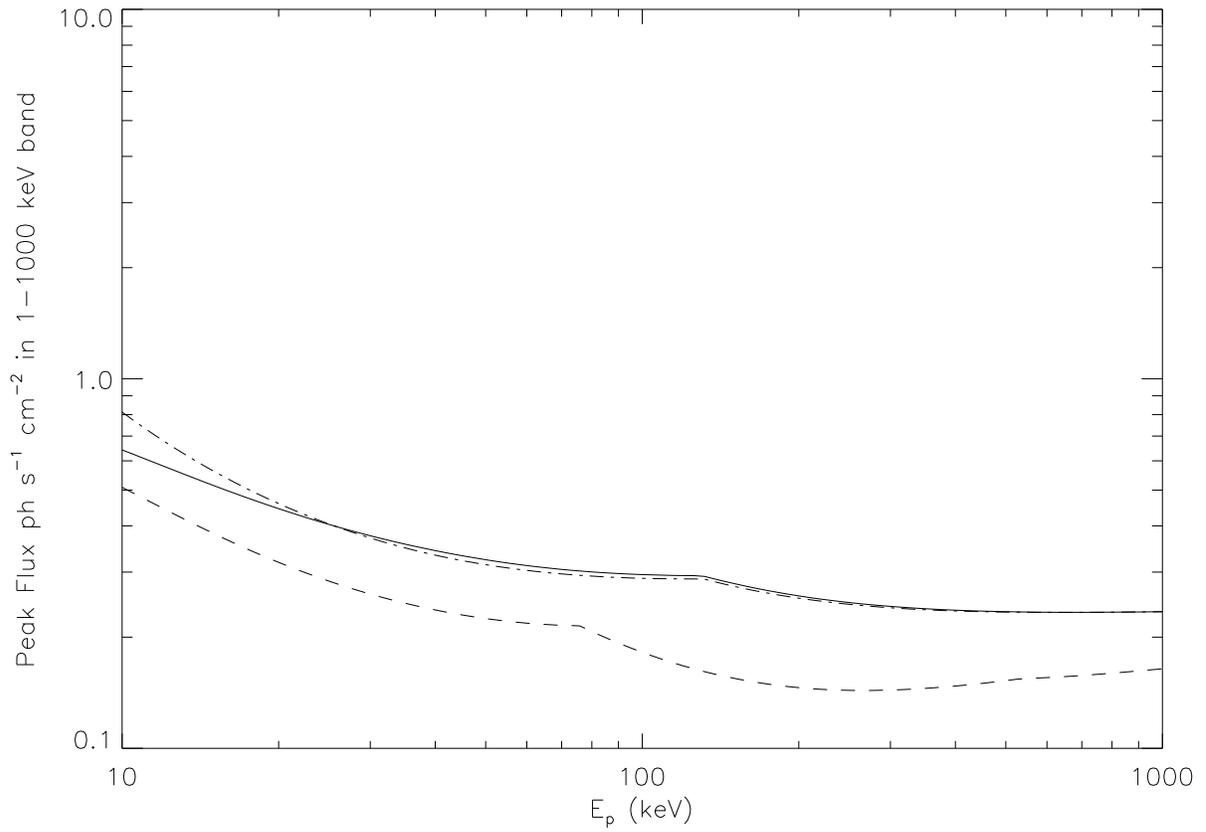}
\caption{Peak flux (1--1000~keV)
threshold of one {\it EXIST} telescope.  Solid
line---$\alpha = -1$, $\beta = -2$; dashed line---$\alpha =
-0.5$, $\beta = -2$; dot-dashed line---$\alpha = -1$,
$\beta = -3$. \label{exist}}
\end{figure}

\begin{figure}
\plotone{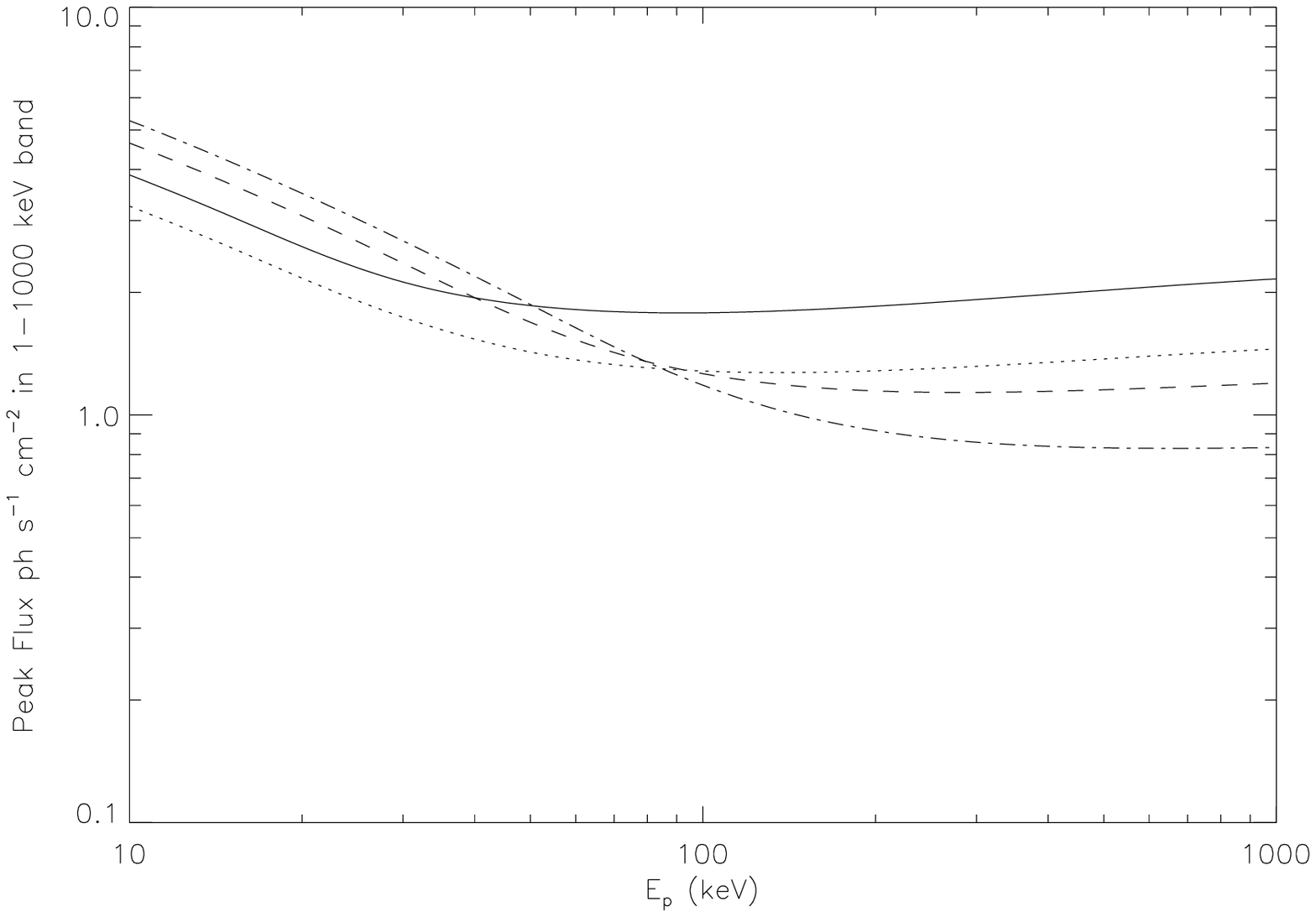}
\caption{Peak flux (1--1000~keV)
threshold for the different Swift energy trigger bands.  The GRB
model with $\alpha = -1$ and $\beta = -2$ is assumed.  The solid
line is for the 15--30~keV band, dotted for 15--50~keV, dashed for
30--75~keV and dot-dashed for 50--150~keV. \label{swift_comp1}}
\end{figure}




\clearpage

\begin{thebibliography}{}

\bibitem[bs]{bw}Amati, L. 1999, thesis at Universit\'a ``La Sapienza'' di Roma
(http://tonno.tesre.bo.cnr.it/~amati/tesi/tesi.html)

\bibitem[y2]{y2}Atteia, J-L., et al. 2002,
in Gamma-Ray Burst and Afterglow Astronomy 2001: A Workshop
Celebrating the First Year of the HETE Mission, eds. R.~Vanderspek
and G.~Ricker, in press [astro-ph/0202515]

\bibitem[yxyx]{yxyx}Band, D. L. 2002, ApJ, 578, 806

\bibitem[yx]{yx}Band, D. L., et al. 1993, ApJ, 413, 281

\bibitem[xb1]{x34}Feroci, M., et al. 1997, Proc. SPIE Conference
(San Diego, 1997), 3114, 186 [astro-ph/9708168]

\bibitem[yy]{yy}Fishman, G. J., et al. 1989, in Proceedings of the
Gamma Ray Observatory Science Workshop, 2-39

\bibitem[al2]{as1}Guidorzi,~C. 2001, thesis at the University of
Ferrara (http://www.fe.infn.it/~guidorzi/doktorthese/)

\bibitem[qw]{qw}Grindlay, J., et al. 2002,
in Gamma-Ray Burst and Afterglow Astronomy 2001: A Workshop
Celebrating the First Year of the HETE Mission, eds.
R.~Vanderspek and G.~Ricker, in press

\bibitem[ay]{ay}Gruber, D. E. 1992, in The X-ray Background.
Collected Papers and Reviews from a Workshop held in Laredo,
Spain, September, 1990, eds. X.~Barcons \& A.~C.~Fabian,
(Cambridge: Cambridge University Press), p.~44

\bibitem[asdf]{asdf}Heise, J., in't Zand, J., Kippen, R.~M., \&
Woods,~P.~M. 2001, in Gamma-Ray Bursts in the Afterglow Era,
17--20 October, 2000, eds. E.~Costa, F.~Frontera, \& J.~Hjorth
(Berlin: Springer), p.~16

\bibitem[aq]{aq}Jager, R., et al. 1997, A\&A Suppl, 125, 557

\bibitem[d77]{d77}Kawai, N., et al. 2002, in Gamma-Ray Burst
and Afterglow Astronomy 2001: A Workshop Celebrating the
First Year of the HETE Mission, eds. R.~Vanderspek and
G.~Ricker, in press

\bibitem[a2]{a2}Mallozzi, R., et al. 1995, ApJ, 454, 597

\bibitem[a3]{a3}Preece, R., et al. 2000, ApJS, 126, 19

\bibitem[a4]{a4}Strohmayer, T., Fenimore, E., Murakami, T.,
\& Yoshida, A. 1998, ApJ, 500, 873

\bibitem[a13]{a13}Sullivan, J. D. 1971, NIM, 95, 5

\bibitem[a7]{a7}von Kienlin, A., et al. 2000, in the 4th Integral
Workshop, Alicante, 2000

\end{thebibliography}
\end{document}